\documentclass{llncs}
\usepackage{makeidx}
\usepackage[dvipsnames]{xcolor}
\usepackage[utf8]{inputenc}
\usepackage{footnote}
\usepackage{listings}
\usepackage{pifont}
\usepackage{graphicx}
\usepackage{url}
\usepackage{pgfplots}
\usepackage{algorithm}
\usepackage{algpseudocode}
\pgfplotsset{width=7cm}
\usepackage{subcaption}
\usepackage{caption}
\DeclareCaptionType{copyrightbox}
\usepackage[multiple]{footmisc}
\usepackage{amsmath}
\usepackage{hyphenat}
\usepackage{hyperref}
\usepackage{bussproofs}
\usepackage{turnstile}
\usepackage{cite}
\hypersetup{
    colorlinks=true,
    linkcolor=blue,
    filecolor=blue,
    urlcolor=blue,
    citecolor = blue
}

\newcommand{\todo}[1]{\textcolor{red}{}}
\makeatletter
\newcommand{\specialterms}[1]{%
  \@for\next:=#1\do
    {\@namedef{specialterm@\detokenize\expandafter{\next}}{}}%
}
\newcommand\term[1]{%
  \@ifundefined{specialterm@\detokenize{#1}}
    {#1}{\emph{#1}\global\expandafter\let\csname specialterm@\detokenize{#1}\endcsname\relax}%
}
\makeatother

\specialterms{C,C++,TU,AST,CFG,CG,DFA,M\'elange,UseDefChecker,BugReportAnalyzer,checker,pass,candidate bugs,whole\hyp{}program pass,libxul,Gecko,xgcc,Function,Module,Branch,Program State,CFG Element,Program Point,Objective C,Whole\hyp{}program LLVM,scan\hyp{}build,Forward Symbolic Execution,Declarations,Declaration Tainting,CallGraph,Basic CallGraph,CWE457,CWE843,CWE194,CWE195,Medium-Severity,AEG,Parfait,Joern,Metal/xgcc,Objective-C,build,checkers,passes}

\newcommand\tw[1]{\term{#1}}
\newcommand\cpp{\term{C++}}

\newcommand{\astr}{\term{AST}}

\newcommand{\pallang}{\term{M\'elange}}
  {\gdef\scalefactor{#1}\begin{center}\proofSkipAmount \leavevmode}%
  {\scalebox{\scalefactor}{\DisplayProof}\proofSkipAmount \end{center} }

\begin{document}

\title{Towards Vulnerability Discovery Using Staged Program Analysis}


\author{Bhargava Shastry\inst{1} \and Fabian Yamaguchi\inst{2} \and
Konrad Rieck\inst{2} \and \\ Jean-Pierre Seifert\inst{1}}
\institute{
Security in Telecommunications, TU Berlin, Germany
\and
Institute of System Security, TU Braunschweig, Germany
}

\maketitle

\begin{abstract}
Eliminating vulnerabilities from low-level code is vital for securing software.
Static analysis is a promising approach for discovering vulnerabilities since it can provide developers early feedback on the code they write.
But, it presents multiple challenges not the least of which is understanding what makes a bug exploitable and conveying this information to the developer.
In this paper, we present the design and implementation of a practical vulnerability assessment framework, called \pallang{}.
\pallang{} performs data and control flow analysis to diagnose potential security bugs, and outputs well\hyp{}formatted bug reports that help developers \textit{understand} and \textit{fix} security bugs.
Based on the intuition that real\hyp{}world vulnerabilities manifest themselves across multiple parts of a program, \pallang{} performs both local and global analyses.
To scale up to large programs, global analysis is demand\hyp{}driven.
Our prototype detects multiple vulnerability classes in \tw{C} and \tw{C++} code including type confusion, and garbage memory reads.
We have evaluated \pallang{} extensively.
Our case studies show that \pallang{} scales up to large codebases such as Chromium, is easy\hyp{}to\hyp{}use, and most importantly, capable of discovering vulnerabilities in real\hyp{}world code.
Our findings indicate that static analysis is a viable reinforcement to the software testing tool set.


\end{abstract}

\section{Introduction}
\label{sec:introduction}
Vulnerabilities in popularly used software are not only detrimental to end\hyp{}user security but can also be hard to identify and fix.
Today's highly inter\hyp{}connected systems have escalated the damage inflicted upon users due to security compromises as well as the cost of fixing vulnerabilities.
To address the threat landscape, software vendors have established mechanisms for software quality assurance and testing.
A prevailing thought is that security bugs identified and fixed early impose lower costs than those identified during the testing phase or in the wild.
Thus, vulnerability re-mediation---the process of identifying and fixing vulnerabilities---is being seen as part of the software development process rather than in isolation~\cite{mssecurity}.

Program analysis provides a practical means to discover security bugs during software development.
Prior approaches to vulnerability discovery using static code analysis have ranged from simple pattern\hyp{}matching to context and path\hyp{}sensitive data\hyp{}flow analysis.
For instance, ITS4~\cite{Viega00}---a vulnerability scanner for \tw{C}/\tw{C++} programs---parses source code and looks up lexical tokens of interest against an existing vulnerability database.
In our initial experiments, the pattern-matching approach employed by ITS4 produced a large number of warnings against modern \tw{C}, and \tw{C++} codebases.
On the contrary, security vulnerabilities are most often, subtle corner cases, and thus rare.
The approach taken by ITS4 is well\hyp{}suited for extremely fast analysis, but the high amount of manual effort required to validate warnings undermines the value of the tool itself.

On the other end of the spectrum, the Clang Static Analyzer~\cite{ClangSA} presents an analytically superior approach for defect discovery.
Precise---context and path sensitive---analysis enables Clang SA to warn only when there is evidence of a bug in a feasible program path.
While precise warnings reduce the burden of manual validation, we find that Clang SA's local inter\hyp{}procedural analysis misses security bugs that span file boundaries.
The omission of bugs that span file boundaries is significant especially for object\hyp{}oriented code\footnote{All the major browsers including Chromium and Firefox are implemented in object\hyp{}oriented code.}, where object implementation and object use are typically in different source files.
A natural solution is to make analysis global.
However, global analysis does not scale up to large programs.

In this paper, we find a middle ground.
We present the design and implementation of \pallang{}, a vulnerability assessment tool for \tw{C} and \tw{C++} programs, that performs both local and global analysis in stages to discover potential vulnerabilities spanning source files.
\pallang{} has been implemented as an extension to the LLVM compiler infrastructure~\cite{llvm}.
To keep analysis scalable, \pallang{} performs computationally expensive analyses locally (within a source file), while performing cheaper analyses globally (across the whole program).
In addition, global analysis is demand\hyp{}driven: It is performed to validate the outcome of local analyses.
To provide good diagnostics, \pallang{} primarily analyzes source code.
It outputs developer\hyp{}friendly bug reports that point out the \textit{exact} position in source code where potential vulnerabilities exist, why they are problematic, and how they can be remedied.

Results from our case studies validate our design decisions.
We find that \pallang{} is capable of highlighting a handful of problematic corner cases, while scaling up to large programs like Chromium, and Firefox.
Since \pallang{} is implemented as an extension to a widely used compiler toolchain (Clang/LLVM), it can be invoked as part of the build process.
Moreover, our current implementation is fast enough to be incorporated into nightly builds\footnote{Regular builds automatically initiated overnight on virtual machine clusters} of two large codebases (MySQL, Chromium), and with further optimizations on the third (Firefox).
In summary, we make the following contributions.
\begin{enumerate}
\item We present the design and implementation of \pallang{}, an extensible program analysis framework.
\item We demonstrate the utility of \pallang{} by employing it to detect multiple classes of vulnerabilities, including garbage reads and incorrectly typed data, that are known to be a common source of exploitable vulnerabilities.
\item We evaluate \pallang{} extensively.
We benchmark \pallang{} against NIST's Juliet benchmark~\cite{juliet} for program analysis tools.
\pallang{} has thus far detected multiple known vulnerabilities in the PHP interpreter, and Chromium codebases, and discovered a new defect in Firefox.
\end{enumerate}

\section{Background: Clang and LLVM}
\label{sec:background}
\pallang{} is anchored in the Clang/LLVM open\hyp{}source compiler toolchain~\cite{llvmweb}, an outcome of pioneering work by Lattner et al.~\cite{llvm}.
In this section, we review components of this toolchain that are at the core of \pallang{}'s design.
While Clang/LLVM is a compiler at heart, it's utility is not limited to code generation/optimization.
Different parts of the compiler front-end (Clang) and back-end (LLVM) are encapsulated into libraries that can be selectively used by client systems depending on their needs.
Thus, the LLVM project lends itself well to multiple compiler-technology-driven use-cases, program analysis being one of them.

We build \pallang{} on top of the analysis infrastructure available within the LLVM project.
This infrastructure mainly comprises the Clang Static Analyzer---a source code analyzer for \emph{C}, \cpp{}, and \tw{Objective-C} programs---and the LLVM analyzer/optimizer framework which permits analysis of LLVM Bitcode.
In the following paragraphs, we describe each of these components briefly.

\subsection{Clang Static Analyzer}
\label{sec:clangsa_overview}

The Clang Static Analyzer (Clang SA) is similar in spirit to \tw{Metal/xgcc}, which its authors classify as a ``Meta-level Compilation'' (MC) framework~\cite{mc, Hallem02}.
The goal of an MC framework is to allow for modular extensions to the compiler that enable checking of domain\hyp{}specific program properties.
Abstractly viewed, an MC framework comprises a set of \emph{checkers} (domain\hyp{}specific analysis procedures) and a compilation system.

\begin{figure}[t]
  \centering
  \includegraphics[width=0.7\textwidth]{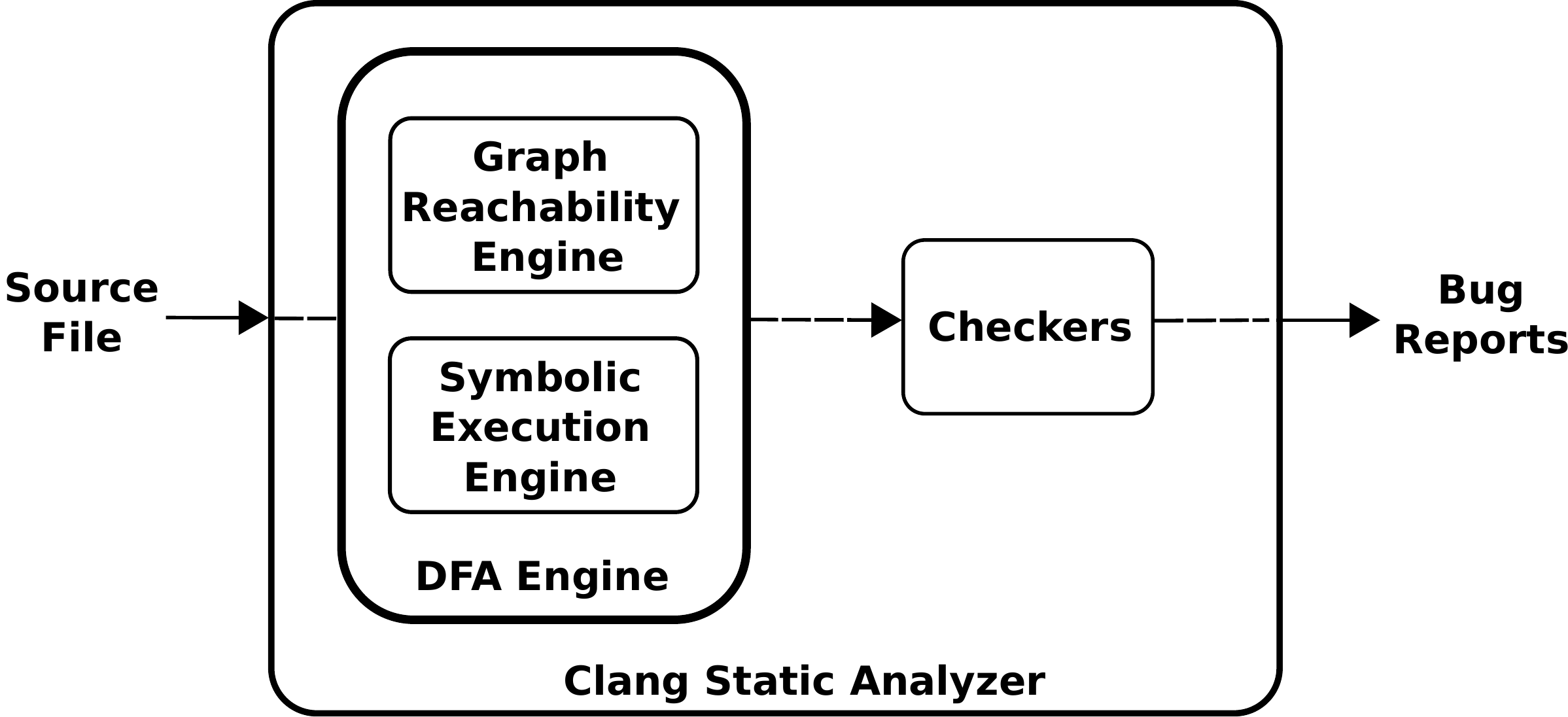}
  \caption{Clang Static Analyzer overview}
  \label{fig:clangsa_overview}
\end{figure}

The division of labor envisioned by Hallem et al.~\cite{Hallem02} is that \tw{checkers} only \emph{encode} the property to check, leaving the mechanics of the actual checking to the compilation system.
The compilation system facilitates checking by providing the necessary analysis infrastructure.
Figure \ref{fig:clangsa_overview} shows how an MC framework is realized in Clang SA.
Source files are parsed and subsequently passed on to the Data-Flow Analysis engine (\tw{DFA} engine), which provides the analysis infrastructure required by checkers.
Checkers encode the program property to be checked and produce bug reports if a violation is found.
Bug reports are then reviewed by a human analyst.

\paragraph*{Data-flow Analysis Engine}
Clang SA performs \emph{Context} and \emph{Path} sensitive inter\hyp{}procedural data-flow analysis.
Context sensitivity means that the analysis preserves the calling context of function calls; path sensitivity means that the analysis explores paths forked by branch statements \emph{independently}.
Context sensitivity is realized in the \emph{Graph Reachability Engine} which implements a namesake algorithm proposed by Reps et al.~\cite{reps95}.
Path sensitivity is implemented in the \emph{Symbolic Execution Engine}.
The symbolic execution engine uses static Forward Symbolic Execution (\tw{FSE})~\cite{fse} to explore program paths in a source file.

\paragraph*{Checkers}
Checkers implement domain\hyp{}specific checks and issue bug reports.
Clang SA contains a default suite of checkers that implement a variety of checks including unsafe API usage, and memory access errors.
More importantly, the checker framework in Clang SA can be used by programmers to add custom checks.
To facilitate customized checks, Clang SA exposes callbacks (as APIs) that \emph{hook} into the DFA engine at pre\hyp{}defined program locations.
Clang SA and its checkers seen together, demonstrate the utility of meta\hyp{}level compilation.

\subsection{LLVM Pass Infrastructure}
\label{sec:llvm_passes}

The LLVM pass infrastructure \cite{llvmweb} provides a modular means to perform analyses and optimizations on an LLVM Intermediate Representation (IR) of a program.
LLVM IR is a typed, yet source\hyp{}language independent representation of a program that facilitates \emph{uniform} analysis of whole\hyp{}programs or whole\hyp{}libraries.

Simply put, an \tw{LLVM Pass} is an operation (procedure invocation) on a unit of LLVM IR code.
The granularity of code operated on can vary from a \tw{Function} to an entire program (\tw{Module} in LLVM parlance).
Passes may be run in sequence, allowing a successive \tw{pass} to reuse information from (or work on a transformation carried out by) preceding \tw{passes}.
The LLVM pass framework provides APIs to tap into source-level meta-data in LLVM IR.
This provides a means to bridge the syntactic gap between source-level and IR-level analyses.
Source literals may be matched against LLVM IR meta-data programmatically.
\pallang{} takes this approach to teach the LLVM pass what a source-level bug report means.

\section{M{\'e}lange}
\label{sec:design}
\begin{figure}[!tbp]
  \centering
  \includegraphics[width=0.7\textwidth]{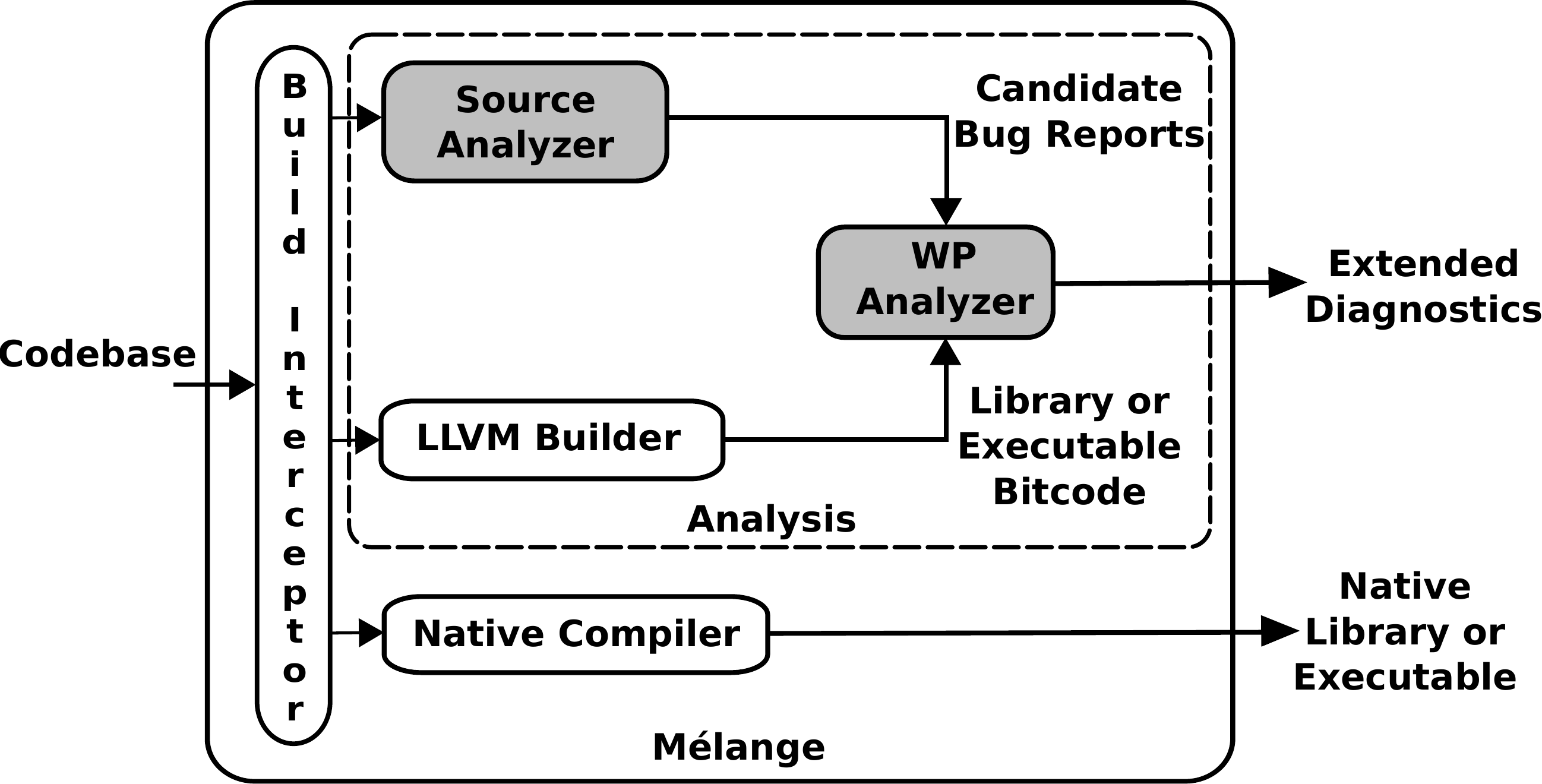}
  \caption{\pallang{} overview}
  \label{fig:pallang_overview}
\end{figure}

Our primary goal is to develop an early warning system for security\hyp{}critical software defects.
We envision \pallang{} as a tool that assists a developer in identifying, and fixing potential security bugs \emph{during} software development.
Figure \ref{fig:pallang_overview} provides an overview of our approach.
\pallang{} comprises four high-level components: the build interceptor, the LLVM builder, the source analyzer, and the Whole\hyp{}Program (WP) analyzer.
We summarize the role of each component in analyzing a program.
Subsequently, we describe them in greater detail.
\begin{enumerate}
\item \emph{Build Interceptor.}
The build interceptor is a program that interposes between the build program (e.g., \emph{GNU-Make}) and the compilation system (e.g., Clang/LLVM).
In \pallang{}, the build interceptor is responsible for \emph{correctly} and \emph{independently} invoking the program builders and the source analyzer. (\S{\ref{sec:analysis_utilities}})
\item \emph{LLVM Builder.}
The LLVM builder is a utility program that assists in generating LLVM Bitcode for \tw{C}, \cpp{}, and \tw{Objective-C} programs.
It mirrors steps taken during native compilation onto LLVM Bitcode generation. (\S{\ref{sec:analysis_utilities}})
\item \emph{Source Analyzer.}
The source analyzer executes domain\hyp{}specific checks on a source file and outputs candidate bug reports that diagnose a potential security bug.
The source analyzer is invoked during the first stage of \pallang{}'s analysis.
We have implemented the source analyzer as a library of checkers that plug into a patched version of Clang SA. (\S{\ref{sec:source_analysis}})
\item \emph{Whole\hyp{}Program Analyzer.}
The WP analyzer examines candidate bug reports (from Step 3), and either provides extended diagnostics for the report or classifies it as a false positive.
The developer is shown only those reports that have extended diagnostics i.e., those not classified as a false positive by the WP analyzer.
We have implemented the WP analyzer in multiple LLVM passes. (\S{\ref{sec:whole_program_analysis}})
\end{enumerate}

\todo{
1. Why are analysis utilities important?
2. How do they interact with the build system?
3. How do they interact with downstream programs?
4. How do they work?
}

\subsection{Analysis Utilities}
\label{sec:analysis_utilities}
Ease\hyp{}of\hyp{}deployment is one of the design goals of \pallang{}.
We want software developers to use our analysis framework in their build environments seamlessly.
The build interceptor and the LLVM builder are \emph{analysis utilities} that help us achieve this goal.
The build interceptor and the LLVM builder facilitate transparent analysis of codebases by \emph{plugging in} \pallang{}'s analyses to an existing build system.
We describe them briefly in the following paragraphs.

\paragraph{Build Interceptor}
Our approach to transparently analyze large software projects hinges on triggering analysis via the build command.
We use an existing build interceptor, \tw{scan\hyp{}build}~\cite{scanbuild}, from the Clang project.
\tw{scan\hyp{}build} is a command\hyp{}line utility that intercepts build commands and invokes the source analyzer in tandem with the compiler.
Since \pallang{}'s WP analysis is targeted at program (LLVM) Bitcode, we instrument \tw{scan\hyp{}build} to not only invoke the source analyzer, but also the LLVM builder.

\paragraph{LLVM Builder}
Generating LLVM Bitcode for program libraries and executables without modifying source code and/or build configuration is a daunting task.
Fortunately, the \tw{Whole\hyp{}program LLVM} (WLLVM)~\cite{wllvm}, an existing open\hyp{}source LLVM builder, solves this problem.
WLLVM is a python\hyp{}based utility that leverages a compiler for generating whole\hyp{}program or whole\hyp{}library LLVM Bitcode.
It can be used as a drop\hyp{}in replacement for a compiler i.e., pointing the builder (e.g., \emph{GNU-Make}) to WLLVM is sufficient.

\todo{
1. Clang SA uses under\hyp{}constrained forward symbolic execution to find defects at compile time
2. While FSE is a powerful technique to find local defects, it is deficient for OO code
3. The concept of encapsulation separates object implementation from use
4. FSE finds bugs when objects are used but not implementation
5. We build on top of Clang SA, leveraging its flow sensitive data-flow analysis engine
6. But we don't really symbolically execute anything, we simply work on object declarations
7. We use object declaration tainting to collect use and def events
8. We consolidate use and def events for each member of all objects encountered during analysis
9. We build a summary of use/def events for each procedure in a source file
10. This allows us to inter-procedurally flag a warning for class members that have a use with a missing def
11. We call such bugs candidate bugs and the summary contained therein bug summary
}

\subsection{Source Analyzer}
\label{sec:source_analysis}
The source analyzer assists \pallang{} in searching for potential bugs in source code.
We build an \textit{event collection} system to aid this search.
Our event collection system is flexible enough to cater to multiple use\hyp{}cases.
It can be employed to detect both traditional taint\hyp{}style vulnerabilities as well as semantic defects.
The event collection system is implemented as a system of taints on \tw{C} and \cpp{} language constructs (\tw{Declarations}).
We call the underlying mechanism \tw{Declaration Tainting} because taints in the proposed event collection system are associated with \tw{AST Declaration} identifiers of \tw{C} and \cpp{} objects.

We write checkers to flag defects.
Checkers have been developed as \emph{clients} of the proposed event collection system.
The division of labor between checkers and the event collection system mirrors the Meta-level Compilation concept: Checkers encode the policy for flagging defects, while the event collection system maintains the state required to perform checks.
We have prototyped this system for flagging garbage (uninitialized) reads\footnote{The algorithm for flagging garbage reads is based on a variation of gen\hyp{}kill sets \cite{genkill}.} of \cpp{} objects, incorrect type casts in PHP interpreter codebase, and other Common Weakness Enumerations (see \S{\ref{sec:evaluation}}).

We demonstrate the utility of the proposed system by using the code snippet shown in Listing \ref{opsem_example} as a running example.
Our aim is to detect uninitialized reads of class members in the example.
The listing encompasses two source files, {\tt foo.cpp} and {\tt main.cpp}, and a header file {\tt foo.h}.
We maintain two sets in the event collection system: the {\tt Def} set containing declaration identifiers for class members that have at least one definition, and the {\tt UseWithoutDef} set containing identifiers for class members that are used (at least once) without a preceding definition.
We maintain an instance of both sets for each function that we analyze in a translation unit i.e., for function $F$, $\Delta{}_{F}$ denotes the analysis summary of $F$ that contains both sets.
The checker decides how the event collection sets are populated.
The logic for populating the {\tt Def} and {\tt UseWithoutDef} sets is simple.
If a program statement in a given function defines a class member for the very first time, we add the class member identifier to the {\tt Def} set of that function's analysis summary.
If a program statement in a given function uses a class member that is absent from the {\tt Def} set, we add the class member identifier to the {\tt UseWithoutDef} set of that function's analysis summary.

\begin{lstlisting}[caption=Running example--The {\tt foo} object does not initialize its class member {\tt foo::x}. The call to {\tt isZero} on Line 22 leads to a garbage read on Line 13., label=opsem_example]
// foo.h
class foo {
public:
        int x;
        foo() {}
        bool isZero();
};

// foo.cpp
#include "foo.h"

bool foo::isZero() { 
  if (!x)
    return true;
}

// main.cpp
#include "foo.h"

int main() {
        foo f;
        if (f.isZero())
          return 0;
        return 1;
}
\end{lstlisting}

In Listing \ref{opsem_example}, when function {\tt foo::isZero} in file {\tt foo.cpp} is being analyzed, the checker adds class member {\tt foo::x} to the \tw{UseWithoutDef} set of $\Delta{}_{foo::isZero}$ after analyzing the branch condition on Line 13.
This is because the checker has not encountered a definition for {\tt foo::x} in the present analysis context.
Subsequently, analysis of the constructor function {\tt foo::foo} does not yield any additions to either the \tw{Def} or \tw{UseWithoutDef} sets.
So $\Delta{}_{foo::foo}$ is empty.
Finally, the checker compares set memberships across analysis contexts.
Since {\tt foo::x} is marked as a use without a valid definition in $\Delta_{foo::isZero}$ and {\tt foo::x} is not a member of the \tw{Def} set in the constructor function's analysis summary ($\Delta{}_{foo::foo}$), the checker classifies the use of Line 13 as a candidate bug.
The checker encodes the proof for the bug in the candidate bug report.
Listing \ref{opsem_report} shows how candidate bug reports are encoded.
The bug report encodes the location and analysis stack corresponding to the potential garbage (uninitialized) read.

The proposed event collection approach has several benefits.
First, by retrofitting simple declaration-based object tainting into Clang SA, we enable \tw{Checkers} to perform analysis based on the proposed taint abstraction.
Due to its general\hyp{}purpose nature, the taint abstraction is useful for discovering other defect types such as null pointer dereferences.
Second, the tainting APIs we expose are opt-in.
They may be used by existing and/or new checkers.
Third, our additions leverage high\hyp{}precision analysis infrastructure already available in Clang SA.
We have implemented the event collection system as a patch to the mainline version of Clang Static Analyzer.
In the next paragraph, we describe how candidate bug reports are analyzed by our whole\hyp{}program analyzer.


\subsection{Whole\hyp{}Program Analyzer}
\label{sec:whole_program_analysis}

\todo{
1. Whole-program analysis is demand-driven
2. Each candidate bug report is analyzed against a larger compilation unit such as a library or executable
3. The analysis is implemented as an LLVM pass
4. Inputs to the pass are encoded as preprocessor directives
5. A script is used to run the pass against all candidate bug reports
6. The analysis itself is composed of several constituent analyses
7. First, call graph for the program is built
8. We use class hierarchy analysis to obtain the call graph
9. Next, we traverse the call graph and keep track of definitions for the uninitialized class member encoded in the candidate bug report
10. Later, we make use of this knowledge to output a call chain in which the candidate bug report might manifest as a garbage read
}

Whole\hyp{}program analysis is demand\hyp{}driven.
Only candidate bug reports are analyzed.
The analysis target is an LLVM Bitcode file of a library or executable.
There are two aspects to WP analysis: Parsing of candidate bug reports to construct a query, and the analysis itself.
We have written a simple python\hyp{}based parser to parse candidate bug reports and construct queries.
The analysis itself is implemented as a set of LLVM passes.
The bug report parser encodes queries as preprocessor directives in a pass header file.
A driver script is used to recompile, and run the pass against all candidate bug reports.

Our whole\hyp{}program analysis routine is composed of a \tw{CallGraph} analysis pass.
We leverage an existing LLVM pass called the \tw{Basic CallGraph} pass to build a whole\hyp{}program call graph.
Since the basic pass misses control flow at indirect call sites, we have implemented additional analyses to improve upon the precision of the basic callgraph.
Foremost among our analyses is Class Hierarchy Analysis (CHA) \cite{cha}.
CHA enables us to devirtualize those dynamically dispatched call sites where we are sure no delegation is possible.
Unfortunately, CHA can only be undertaken in scenarios where no new class hierarchies are introduced.
In scenarios where CHA is not applicable, we examine call instructions to resolve as many forms of indirect call sites as possible.
Our prototype resolves aliases of global functions, function casts etc.

Once program call graph has been obtained, we perform a domain\hyp{}specific WP analysis.
For instance, to validate garbage reads, the pass inspects loads and store to the buggy program variable or object.
In our running example (Listing \ref{opsem_example}), loads and stores to the {\tt foo::x} class member indicated in candidate bug report (Listing \ref{opsem_report}) are tracked by the WP garbage read pass.
To this end, the program call graph is traversed to check if a load of {\tt foo::x} does not have a matching store.
If all loads have a matching store, the candidate bug report is classified as a false positive.
Otherwise, program call-chains in which a load from {\tt foo::x} does not have a matching store are displayed to the analyst in the whole\hyp{}program bug report (Listing \ref{opsem_report}).

\begin{lstlisting}[numbers=none, caption=Candidate bug report (top) and whole\hyp{}program bug report (bottom) for garbage read in the running example shown in Listing \ref{opsem_example}., label=opsem_report]
// Source-level bug report
// report-e6ed9c.html
...
Local Path to Bug: foo::x->_ZN3foo6isZeroEv

Annotated Source Code
foo.cpp:4:6: warning: Potentially uninitialized object field
 if (!x)
      ^
1 warning generated.

// Whole-program bug report
---------- report-e6ed9c.html ---------
[+] Parsing bug report report-e6ed9c.html
[+] Writing queries into LLVM pass header file
[+] Recompiling LLVM pass
[+] Running LLVM BugReportAnalyzer pass against main
---------------------------------------
Candidate callchain is: 

foo::isZero()
main
-----------------------
\end{lstlisting}

\section{Evaluation}
\label{sec:evaluation}
We have evaluated \pallang{} against both static analysis benchmarks and real\hyp{}world code.
To gauge \pallang{}'s utility, we have also tested it against known defects and vulnerabilities.
Our evaluation seeks to answer the following questions:
\begin{itemize}
\item What is the effort required to use \pallang{} in an existing build system? (\S{\ref{sec:deployability}})
\item How does \pallang{} perform against static analysis benchmarks? (\S{\ref{sec:nist_juliet}})
\item How does \pallang{} fare against known security vulnerabilities? (\S{\ref{sec:known_vulns}})
\item What is the analysis run\hyp{}time and effectiveness of \pallang{} against large well\hyp{}tested codebases? (\S{\ref{sec:case_study}})
\end{itemize}

\subsection{Deployability}
\label{sec:deployability}

Ease\hyp{}of\hyp{}deployment is one of the design goals of \pallang{}.
Build interposition allows us to analyze codebases as is, without modifying build configuration and/or source code.
We have deployed \pallang{} in an Amazon {\tt compute} instance where codebases with different build systems have been analyzed (see \S{\ref{sec:case_study}}).
Another benefit of build system integration is incremental analysis.
Only the very first build of a codebase incurs the cost of end\hyp{}to\hyp{}end analysis; subsequent analyses are incremental.
While incremental analysis can be used in conjunction with daily builds, full analysis can be coupled with nightly builds and initiated on virtual machine clusters.


\subsection{NIST Benchmarks}
\label{sec:nist_juliet}

We used static analysis benchmarks released under NIST's SAMATE project \cite{samate} for benchmarking \pallang{}'s detection rates.
In particular, the Juliet \tw{C}/\cpp{} test suite (version 1.2) \cite{juliet} was used to measure true and false positive detection rates for defects spread across multiple categories.
The Juliet suite comprises test sets for multiple defect types.
Each test set contains test cases for a specific Common Weakness Enumeration (CWE) \cite{seven}.
The CWE system assigns identifiers for common classes of software weaknesses that are known to lead to exploitable vulnerabilities.
We implemented \pallang{} \tw{checkers} and \tw{passes} for the following CWE categories: \tw{CWE457} (Garbage or uninitialized read), \tw{CWE843} (Type confusion), \tw{CWE194} (Unexpected Sign Extension), and \tw{CWE195} (Signed to Unsigned Conversion Error).
With the exception of \tw{CWE457}, the listed CWEs have received scant attention from static analysis tools.
For instance, type confusion (\tw{CWE843}) is an emerging attack vector \cite{tcv} for exploiting popular applications.

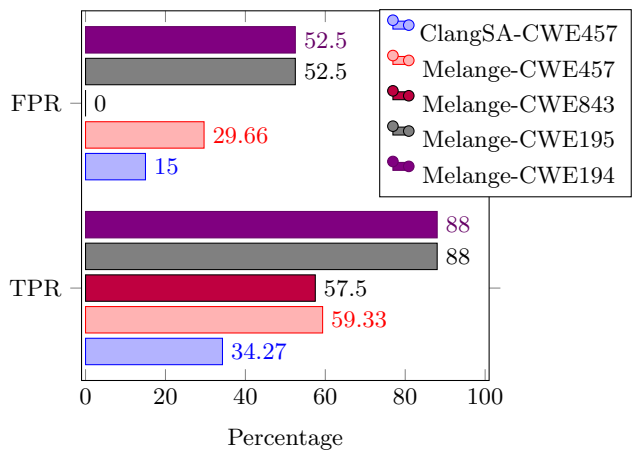
\begin{figure}[t]
\centering
\begin{tikzpicture}
\begin{axis}[
        xbar,
        height=6.5cm,
        xmin=0, xmax=100,
        symbolic y coords={TPR, FPR},
        ytick=data,
 	xlabel= Percentage,
 	legend style={at={(1.35,.75)},
 	anchor=east,legend columns=1},
        legend image post style={mark=*},
        enlargelimits=0.15,
        nodes near coords,
        enlarge y limits=0.5,
        enlarge x limits=0.01,
        nodes near coords align={horizontal},
]
\addplot coordinates {(34.27,TPR) (15,FPR)};
\addplot coordinates {(59.33,TPR) (29.66,FPR)};
\addplot [fill=purple] coordinates {(57.5,TPR) (0,FPR)};
\addplot coordinates {(88,TPR) (52.5,FPR)};
\addplot coordinates {(88,TPR) (52.5,FPR)};
\legend{ClangSA-CWE457,Melange-CWE457,Melange-CWE843,Melange-CWE195,Melange-CWE194}
\end{axis}
\end{tikzpicture}
  \caption{Juliet test suite: True Positive Rate (TPR) and False Positive Rate (FPR) for \pallang{}, and Clang Static Analyzer. Clang SA supports CWE457 only.}
  \label{fig:juliet_eval}
\end{figure}

Figure \ref{fig:juliet_eval} summarizes the True/False Positive Rates (TPRs/FPRs) for Clang SA and \pallang{} for the chosen CWE benchmarks.
Currently, Clang SA only supports \tw{CWE457}.
Comparing reports from Clang SA and \pallang{} for the CWE457 test set, we find that the former errs on the side of precision (fewer false positives), while the latter errs on the side of caution (fewer false negatives).
For the chosen CWE benchmarks, \pallang{} attains a true\hyp{}positive rate between 57--88 \%, and thus, it is capable of spotting over half of the bugs in the test suite.

\pallang{}'s staggered analysis approach allows it to present both source file wide and program wide diagnostics (see Figure \ref{fig:pallang_br}).
In contrast, Clang SA's diagnostics are restricted to a single source file.
Often, the call stack information presented in \pallang{}'s extended diagnostics has speeded up manual validation of bug reports.

\subsection{Detection of Known Vulnerabilities}
\label{sec:known_vulns}

We tested five known type\hyp{}confusion vulnerabilities in the PHP interpreter with \pallang{}.
All of the tested flaws are taint\hyp{}style vulnerabilities: An attacker\hyp{}controlled input is passed to a security\hyp{}sensitive function call that wrongly interprets the input's type.
Ultimately all these vulnerabilities result in invalid memory accesses that can be leveraged by an attacker for arbitrary code execution or information disclosure.
We wrote a checker for detecting multiple instances of this vulnerability type in the PHP interpreter codebase.
For patched vulnerabilities, testing was carried out on unpatched versions of the codebase.
\pallang{} successfully flagged all known vulnerabilities.
The first five entries of Table \ref{table:known_vulns} summarize \pallang{}'s findings.
Three of the five vulnerabilities have been assigned Common Vulnerabilities and Exposures (CVE) identifiers by the MITRE Corporation.
Reporters of {\tt CVE-2014-3515}, {\tt CVE-2015-4147}, and PHP report ID {\tt 73245} have received bug bounties totaling \$5500 by the Internet Bug Bounty Panel \cite{ibb}.

In addition, we ran our checker against a recent PHP release candidate (PHP 7.0 RC7) released on 12th November, 2015.
Thus far, \pallang{} has drawn attention to PHP sub\hyp{}systems where a similar vulnerability may exist.
While we haven't been able to verify if these are exploitable, this exercise demonstrates \pallang{}'s utility in bringing attention to multiple instances of a software flaw in a large codebase that is under active development.

\begin{table}[t]
\centering
\begin{tabular}{ p{2cm} p{3cm} p{2cm} p{2.5cm} p{2cm} }
    \hline
    Codebase & CVE ID (Rating) & Bug ID & Vulnerability & Known/New \\ \hline
    PHP & CVE-2015-4147 & 69085 \cite{P69085} & Type-confusion & Known \\
    PHP & CVE-2015-4148 & 69085 \cite{P69085} & Type-confusion & Known \\
    PHP & CVE-2014-3515 & 67492 \cite{P67492} & Type-confusion & Known \\
    PHP & Unassigned & 73245 \cite{P73245} & Type-confusion & Known \\
    PHP & Unassigned & 69152 \cite{P69152} & Type-confusion & Known \\
    Chromium & (Medium-Severity) & 411177 \cite{C411177} & Garbage read & Known \\
    Chromium & None & 436035 \cite{C436035} & Garbage read &  Known \\
    Firefox & None & 1168091 \cite{F1168091} & Garbage read & New \\ \hline
\end{tabular}
    \caption[Table]{Detection summary of \pallang{} against production codebases. \pallang{} has confirmed known vulnerabilities and flagged a new defect in Firefox. Listed Chromium and Firefox bugs are not known to be exploitable. Chromium bug 411177 is classified as a \tw{Medium-Severity} bug in Google's internal bug tracker.}
    \label{table:known_vulns}
\end{table}

\subsection{Case Studies}
\label{sec:case_study}
To further investigate the practical utility of \pallang{}, we conducted case studies with three popular open\hyp{}source projects, namely, Chromium, Firefox, and MySQL.
We focused on detecting garbage reads only.
In the following paragraphs, we present results from our case studies emphasizing analysis effectiveness, and analysis run\hyp{}time.

\textbf{Software Versions:}
Evaluation was carried out for Chromium version 38 (dated August 2014), for Firefox revision 244208 (May 2015), and for MySQL version 5.7.7 (April 2015).

\textbf{Evaluation Setup:}
Analysis was performed in an Amazon compute instance running Ubuntu 14.04 and provisioned with 36 virtual (Intel Xeon E5-2666 v3) CPUs clocked at 2.6 GHz, 60 GB of RAM, and 100 GB of SSD-based storage.

\subsubsection{Effectiveness}
\label{sec:cs_effectiveness}

\begin{figure}[t]
  \centering
  \includegraphics[width=0.7\textwidth]{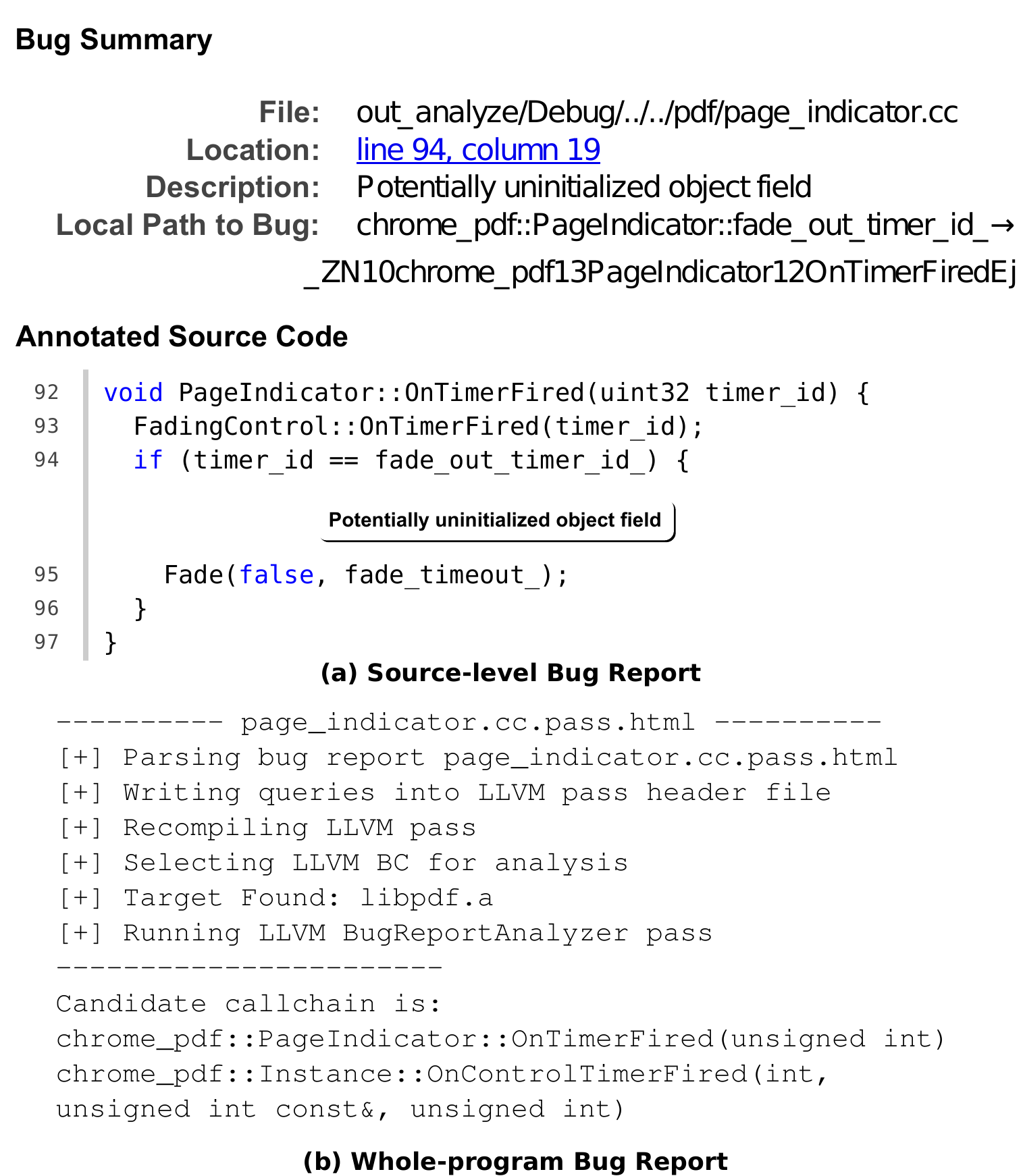}
  \caption{\pallang{} bug report for Chromium bug 411177.}
  \label{fig:pallang_br}
\end{figure}

\paragraph*{True Positives}
\label{true_positives}
Our prototype flagged 3 confirmed defects in Chromium, and Firefox, including a new defect in the latter (see bottom three entries of Table \ref{table:known_vulns}).
Defects found by our prototype in MySQL codebase have been reported upstream and are being triaged.
Figure \ref{fig:pallang_br} shows \pallang{}'s bug report for a garbage read in the pdf library shipped with Chromium v38.
The source-level bug report (Figure \ref{fig:pallang_br}a) shows the line of code that was buggy.
WP analyzer's bug report (Figure \ref{fig:pallang_br}b) shows candidate call chains in the libpdf library in which the uninitialized read may manifest.

We have manually validated the veracity of all bug reports generated by \pallang{} through source code audits.
For each bug report, we verified if the data\hyp{}flow and control\hyp{}flow information conveyed in the report tallied with program semantics.
We classified only those defects that passed our audit as true positives.
Additionally, for the Chromium true positives, we matched \pallang{}'s findings with reports~\cite{C411177,C436035} generated by MemorySanitizer~\cite{tsanmsan}, a dynamic program analysis tool from Google.
The new defect discovered in Firefox was reported upstream~\cite{F1168091}.
Our evaluation demonstrates that \pallang{} can complement dynamic program analysis tools in use today.

\paragraph*{False Positives}
\label{false_positives}

Broadly, we encounter two kinds of false positives; those that are due to imprecision in \pallang{}'s data\hyp{}flow analysis, and those due to imprecision in its control\hyp{}flow analysis.
In the following paragraphs, we describe one example of each kind of false positive.

\textbf{Data\hyp{}flow imprecision:}
\pallang{}'s analyses for flagging garbage reads lack sophisticated alias analysis.
For instance, initialization of \cpp{} objects passed\hyp{}by\hyp{}reference is missed.
Listing \ref{alias} shows a code snippet borrowed from the Firefox codebase that illustrates this category of false positives.

When {\tt AltSvcMapping} object is constructed (see Line 2 of Listing \ref{alias}), one of its class members {\tt mHttps} is passed by reference to the callee function {\tt SchemeIsHTTPS}.
The callee function {\tt SchemeIsHTTPS} initializes {\tt mHttps} via its alias ({\tt outIsHTTPS}).
\pallang{}'s garbage read checker misses the aliased store and incorrectly flags the use of class member {\tt mHttps} on Line 8 as a candidate bug.
\pallang{}'s garbage read pass, on its part, tries to taint all functions that store to {\tt mHttps}.
Since the store to {\tt mHttps} happens via an alias, the pass also misses the store and outputs a legitimate control\hyp{}flow sequence in its WP bug report.

\textbf{Control\hyp{}flow imprecision:}
\pallang{}'s WP analyzer misses control\hyp{}flow information at indirect call sites e.g., virtual function invocations.
Thus, class members that are initialized in a call sequence comprising an indirect function call are not registered by \pallang{}'s garbage read pass.
While resolving all indirect call sites in large programs is impossible, we employ best\hyp{}effort devirtualization techniques such as Rapid Type Analysis~\cite{Bacon96} to improve \pallang{}'s control\hyp{}flow precision.

\begin{lstlisting}[caption=Code snippet involving an aliased definition that caused a false positive in \pallang{}., label=alias]
AltSvcMapping::AltSvcMapping(...) {
  <@\textcolor{ForestGreen}{if (NS\_FAILED(SchemeIsHTTPS(originScheme, mHttps))) \{}@>
    ...
  }
}
void AltSvcMapping::GetConnectionInfo(...) {
  // ci is an object on the stack
  <@\textcolor{ForestGreen}{ci-$>$SetInsecureScheme(!mHttps);}@>
  ...
}
static nsresult SchemeIsHTTPS(const nsACString &originScheme, bool &outIsHTTPS)
{
  <@\textcolor{ForestGreen}{outIsHTTPS $=$ originScheme.Equals(NS\_LITERAL\_CSTRING("https"));}@>
  ...
}
\end{lstlisting}

\begin{table}[t]
\centering
\begin{tabular}{ c  c  c  c  c  c  c  c }
    \hline
    Codebase & Build Time & \multicolumn{4}{c}{Analysis Run\hyp{}time{$^{*}$}} & \multicolumn{2}{c}{Bug Reports} \\ \hline
    & $N_{t}$ & $SA_{x}$ & $WPA_{x}$ & $TA_{x}$ & $WPAvg_{t}$ & Total & True Positives \\ \hline
    Chromium & 18m20s & 29.09 & 15.49 & 44.58 & 7.5s & 12 & 2 \\ \hline
    Firefox & 41m25s & 3.38 & 39.31 & 42.69 & 13m35s & 16 & 1 \\ \hline
    MySQL & 8m15s & 9.26 & 21.24 & 30.50 & 2m26s & 32 & - \\ \hline
\end{tabular}
    \caption*{
    $^{*}$\scriptsize{All terms except $WPAvg$ are normalized to native compilation time}
    }
    \caption[Table]{\pallang{}: Analysis summary for large open-source projects. True positives for MySQL have been left out since we are awaiting confirmation from its developers.}
    \label{table:os_projects}
\end{table}

The last two columns of Table \ref{table:os_projects} present a summary of \pallang{}'s bug reports for Chromium, Firefox, and MySQL projects once both stages of analysis have been completed.
We find that \pallang{}'s two\hyp{}stage analysis pipeline is very effective at filtering through a handful of bug reports that merit attention.
For instance, \pallang{}'s analyses output only twelve bug reports for Chromium, a codebase that spans over 14 million lines of code.
Although \pallang{}'s true positive rate is low in our case studies, the corner cases it has pointed out, notwithstanding the confirmed bugs it has flagged, is encouraging.
Given that we evaluated \pallang{} against well\hyp{}tested production code, the fact that it could point out three confirmed defects in the Chromium and Firefox codebases is a promising result.
We plan to make our tool production\hyp{}ready by incorporating insights gained from our case studies.
Next, we discuss \pallang{}'s analysis run\hyp{}time.

\subsubsection{Analysis Run\hyp{}Time}
\label{sec:analysis_overhead}


We completed end\hyp{}to\hyp{}analysis of Chromium, Firefox, and MySQL codebases---all of which have millions of lines of code---in under 48 hours.
Of these, MySQL, and Chromium were analyzed in a little over 4 hours, and 13 hours respectively.
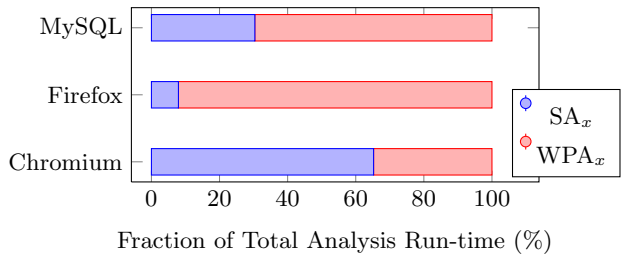
\begin{figure}[!t]
\centering
\begin{tikzpicture}
\begin{axis}[
        xbar stacked,
        height = 3.9cm,
        symbolic y coords={Chromium,Firefox,MySQL},
 	xlabel=Fraction of Total Analysis Run\hyp{}time (\%),
 	legend style={at={(1.2,0.28)},
 	anchor=east,legend columns=1},
        legend image post style={mark=*},
        enlargelimits=0.15,
        nodes near coords align={horizontal},
]
\addplot coordinates {(65.24,Chromium) (7.93,Firefox) (30.37,MySQL)};
\addplot coordinates {(34.76,Chromium) (92.07,Firefox) (69.63,MySQL)};
\addlegendentry{SA$_{x}$}
\addlegendentry{WPA$_{x}$}
\end{axis}
\end{tikzpicture}
  \caption{For each codebase, its source and whole\hyp{}program analysis run\hyp{}times are shown as fractions (in \%) of \pallang{}'s total analysis run\hyp{}time.}
  \label{fig:overheads}
\end{figure}
Table~\ref{table:os_projects} summarizes \pallang{}'s run\hyp{}time for our case studies.
We have presented the analysis run\hyp{}time of a codebase relative (normalized) to its build time, $N_{t}$.
For instance, a normalized analysis run\hyp{}time of 30 for a codebase indicates that the time taken to analyze the codebase is 30\textbf{x} longer than its build time.
All normalized run\hyp{}times are denoted with the $x$ subscript.
Normalized source analysis time, WP analysis time, and total analysis time of \pallang{} are denoted as $SA_{x}$, $WPA_{x}$, and $TA_{x}$ respectively.
The term $WPAvg_{t}$ denotes the average time (not normalized) taken by \pallang{}'s WP analyzer to analyze a single candidate bug report.

Figure~\ref{fig:overheads} shows source and WP analysis run\hyp{}times for a codebase as a fraction (in percentage terms) of \pallang{}'s total analysis run\hyp{}time.
Owing to Chromium's modular build system, we could localize a source defect to a small\hyp{}sized library.
The average size of program analyzed for Chromium (1.8MB) was much lower compared to MySQL (150MB), and Firefox (1.1GB).
As a consequence, the WP analysis run\hyp{}times for Firefox, and MySQL are relatively high.
While our foremost priority while prototyping \pallang{} has been functional effectiveness, our implementation leaves significant room for optimizations that will help bring down \pallang{}'s end\hyp{}to\hyp{}end analysis run\hyp{}time.

\subsection{Limitations}
\label{sec:discussion}

\paragraph{Approach Limitations}
By design, \pallang{} requires two analysis procedures at different code abstractions for a given defect type.
We depend on programmer\hyp{}written analysis routines to scale out to multiple defect types.
Two actualities lend credence to our approach: First, analysis infrastructure required to carry out extended analyses is already available and its use is well\hyp{}documented.
This has assisted us in prototyping \pallang{} for four different CWEs.
Second, the complexity of analysis routines is many times lower than the program under analysis.
Our analysis procedures span $2,598$ lines of code in total, while our largest analysis target (Chromium) has over $14$ million lines of \cpp{} code.

While \pallang{} provides precise diagnostics for security bugs it has discovered, manual validation of bug reports is still required.
Given that software routinely undergoes manual review during development, our tool does not introduce an additional requirement.
Rather, \pallang{}'s diagnostics bring attention to problematic corner cases in source code.
The manual validation process of \pallang{}'s bug reports may be streamlined by subsuming our tool under existing software development processes (e.g., nightly builds, continuous integration).

\paragraph{Implementation Limitations}
\pallang{}'s WP analysis is path and context insensitive.
This makes \pallang{}'s whole\hyp{}program analyzer imprecise and prone to issuing false warnings.
To counter imprecision, we can augment our WP analyzer with additional analyses.
Specifically, more powerful alias analysis and aggressive devirtualization algorithms will help prune false positives further.
One approach to counter existing imprecision is to employ a ranking mechanism for bug reports (e.g., Z-Ranking \cite{zranking}).



\section{Related Work}
\label{sec:related_work}
Program analysis research has garnered attention since the late 70s.
Lint~\cite{lint}, a \tw{C} program checker developed at Bell Labs in 1977, was one of the first program analysis tools to be developed.
Lint's primary goal was to check ``portability, style, and efficiency'' of programs.
Ever since, the demands from a program checker have grown as new programming paradigms have been invented and programs have increased in complexity.
This has contributed to the development of many commercial~\cite{coverity,codesonar,fortify}, closed-source~\cite{parfait}, free~\cite{havoc}, and open source~\cite{ClangSA, asan, tsanmsan, slam, blast, foster03, cqualpp, klee, aeg, vulnextrapolation} tools.
Broadly, these tools are based on \textit{Model Checking}~\cite{slam, blast}, \textit{Theorem Proving}~\cite{havoc}, \textit{Static Program Analysis}~\cite{ClangSA, fortify, coverity, codesonar, parfait, vulnextrapolation}, \textit{Dynamic Analysis}~\cite{valgrind, asan, tsanmsan, klee}, or are hybrid systems such as AEG~\cite{aeg}.
In the following paragraphs, we comment on related work that is close in spirit to \pallang{}.

\paragraph*{Program Instrumentation}
Traditionally, memory access bugs have been found by fuzz testing (or fuzzing) instrumented programs.
The instrumentation takes care of tracking the state of program memory and adds run-time checks before memory accesses are made.
Instrumentation is done either during run time (as in Valgrind~\cite{valgrind}), or at compile time (as in AddressSanitizer or ASan~\cite{asan}).
Compile\hyp{}time instrumentation has been preferred lately due to the poor performance of tools that employ run\hyp{}time instrumentation.

While sanitizer tools such as ASan, and MemorySanitizer (MSan) are expected to have a zero false positive rate, practical difficulties, such as uninstrumented code in an external library, lead to false positives in practice.
Thus, even run\hyp{}time tools do not eliminate the need for manual validation of bug reports.
To guarantee absence of uninitialized memory, MSan needs to monitor each and every load from/store to memory.
This all\hyp{}or\hyp{}nothing philosophy poses yet another problem.
Uninstrumented code in pre-compiled libraries (such as the \tw{C++} standard library) used by the program will invariably lead to false program crashes.
Until these false crashes are rectified---either by instrumenting the code where the crash happens or by asking the tool to suppress the warning---the sanitizer tool is rendered unusable.
Thus, use of MSan impinges on instrumentation of each and every line of code that is directly or indirectly executed by the program or maintenance of a blacklist file that records known false positives.
Unlike MSan, not having access to library source code only lowers \pallang{}'s analysis accuracy, but does not impede analysis itself.
Having said that, \pallang{} will benefit from a mechanism to suppress known false positives.
Overall, we believe that dynamic tools are invaluable for vulnerability assessment, and that a tool such as ours can complement them well.

\paragraph*{Symbolic Execution}
Symbolic execution has been used to find bugs in programs, or to generate test cases with improved code coverage.
KLEE~\cite{klee}, Clang SA~\cite{ClangSA}, and AEG~\cite{aeg} use different flavors of forward symbolic execution for their own end.
As the program (symbolically) executes, constraints on program paths (path predicates) are maintained.
Satisfiability queries on path predicates are used to prune infeasible program paths.
Unlike KLEE and AEG, symbolic execution in Clang SA is done locally and hences scales up to large codebases.
Anecdotal evidence suggests that KLEE and AEG don't scale up to large programs~\cite{acadsucks}.
To the best of our knowledge, KLEE has not been evaluated against even medium\hyp{}sized codebases let alone large codebases such as Firefox and Chromium.

\paragraph*{Static Analysis}
Parfait~\cite{parfait} employs an analysis strategy that is similar in spirit to ours.
It employs multiple stages of analysis, where each successive stage is more precise than the preceding stage.
Parfait has been used for finding buffer overflows in C programs.
In contrast, we have evaluated \pallang{} against multiple vulnerability classes.
\pallang{}'s effectiveness in detecting multiple CWEs validates the generality of its design.
In addition, \pallang{} has fared well against multiple code paradigms: both legacy \tw{C} programs and modern object\hyp{}oriented code.

Like Yamaguchi et al.~\cite{vulnextrapolation}, our goal is to empower developers in finding multiple instances of a known defect.
However, the approach we take is different.
Yamaguchi et al.~\cite{vulnextrapolation}, use structural traits in a program's \astr{} representation to drive a Machine Learning (ML) phase.
The ML phase \textit{extrapolates} traits of known vulnerabilities in a codebase, obtaining matches that are similar in structure to the vulnerability.
CQUAL~\cite{foster03}, and CQual$++$~\cite{cqualpp}, are flow-insensitive data\hyp{}flow analysis frameworks for \tw{C} and \cpp{} languages respectively.
Oink performs whole-program data-flow analysis on the back of Elsa, a \cpp{} parser, and Cqual$++$.
Data-flow analysis is based on type qualifiers.
Our approach has two advantages over Cqual$++$.
We use a production compiler for parsing \cpp{} code that has a much better success rate at parsing advanced \cpp{} code than a custom parser such as Elsa.
Second, our source-level analysis is both flow and path sensitive while, in CQual$++$, it is not.

Finally, Clang Static Analyzer borrows ideas from several publications including (but not limited to)~\cite{reps95, Hallem02}.
Inter-procedural context-sensitive analysis in Clang SA is based on the graph reachability algorithm proposed by Reps et al.~\cite{reps95}.
Clang SA is also similar in spirit to Metal/xgcc~\cite{Hallem02}.

\section{Conclusion}
\label{sec:conclusion}
We have developed \pallang{}, a static analysis tool for helping fix security\hyp{}critical defects in open\hyp{}source software.
Our tool is premised on the intuition that vulnerability search necessitates multi-pronged analysis.
We anchor \pallang{} in the Clang/LLVM compiler toolchain, leveraging source analysis to build a corpus of defects, and whole\hyp{}program analysis to filter the corpus.
We have shown that our approach is capable of identifying defects and vulnerabilities in open\hyp{}source projects, the largest of which---Chromium---spans over $14$ million lines of code.
We have also demonstrated that \pallang{}'s analyses are viable by empirically evaluating its run\hyp{}time in an EC2 instance.

Since \pallang{} is easy to deploy in existing software development environments, programmers can receive early feedback on the code they write.
Furthermore, our analysis framework is extensible via compiler plug-ins.
This enables programmers to use \pallang{} to implement domain\hyp{}specific security checks.
Thus, \pallang{} complements traditional software testing tools such as fuzzers.
Ultimately, our aim is to use the proposed system to help fix vulnerabilities in open\hyp{}source software at an early stage.

\subsubsection*{Acknowledgments}
This work was supported by the following grants: 317888 (project NEMESYS), 10043385 (project Enzevalos), and RI 2468/1-1 (project DEVIL).
Authors would like to thank colleagues at SecT and Daniel Defreez for valuable feedback on a draft of this paper, and Janis Danisevskis for discussions on the \cpp{} standard and occasional code reviews.

\bibliographystyle{splncs03}
\bibliography{master}

\end{document}